\pdfoutput=1

\documentclass[journal,twoside]{IEEEtran}
\usepackage{cite} 
\usepackage{subfigure}
\usepackage{subeqn}
\usepackage{booktabs}
\usepackage{multicol}
\usepackage{lscape,multirow}

\usepackage{ifthen}
\ifx\pdfoutput\undefined 
  \usepackage[dvips]{graphicx}
  \usepackage[ps2pdf]{hyperref}
  \hypersetup{ pdfcreator  = {LaTeX with hyperref package},
               pdfproducer = {dvips + ps2pdf} }
  \usepackage[dvipsnames,dvips,rgb]{xcolor}		
\else 
  \usepackage[pdftex]{graphicx}
  \usepackage[pdftex]{hyperref}
  \usepackage[pdftex,usenames,dvipsnames]{xcolor}
\fi

\hypersetup{
colorlinks=false,
pdfauthor   = {Withayachumnankul~et~al.},
pdftitle    = {T-ray Sensing and Imaging},
pdfsubject  = {},
pdfkeywords = {},
pdfview     = {FitH}
}

\begin{document}

\title{Numerical Removal of Water-Vapor Effects from THz-TDS Measurements}
\author{Withawat~Withayachumnankul,~\IEEEmembership{}
        Bernd~M.~Fischer,~\IEEEmembership{}
        Samuel~P.~Mickan,~\IEEEmembership{Member,~IEEE,}
        and~Derek~Abbott,~\IEEEmembership{Fellow,~IEEE}
\thanks{Manuscript received August 9, 2007; revised ..., 2007.
        This program was supported by The Australian Research Council (ARC), the Sir Ross and Sir Keith Smith Fund, the Defence Science and Technology Organisation (DSTO), and NHEW P/L.}
\thanks{The authors are with the Centre for Biomedical Engineering (CBME) and the School of Electrical \& Electronic Engineering, The University of Adelaide, SA 5005, Australia.}
\thanks{S.~P.~Mickan is also with Davies Collison Cave, 1 Nicholson Street, Melbourne, VIC 3000, Australia.} \thanks{Corresponding author: W. W. (\mbox{withawat@eleceng.adelaide.edu.au}).}
}

\markboth{}{Withayachumnankul \MakeLowercase{\textit{et al.}}: Numerical removal of water-vapor effects from THz-TDS measurements}

\maketitle

\begin{abstract}
One source of disturbance in a pulsed T-ray signal is attributed to ambient water vapor. Water molecules in the gas phase selectively absorb T-rays at discrete frequencies corresponding to their molecular rotational transitions. This results in prominent resonances spread over the T-ray spectrum, and in the time domain the T-ray signal is observed as fluctuations after the main pulse. These effects are generally undesired, since they may mask critical spectroscopic data. So, ambient water vapor is commonly removed from the T-ray path by using a closed chamber during the measurement. Yet, in some applications a closed chamber is not applicable. This situation, therefore, motivates the need for another method to reduce these unwanted artifacts. This paper presents a study on a computational means to address the problem. Initially, a complex frequency response of water vapor is modeled from a spectroscopic catalog. Using a deconvolution technique, together with fine tuning of the strength of each resonance, parts of the water-vapor response are removed from a measured T-ray signal, with minimal signal distortion.
\end{abstract}

\IEEEpeerreviewmaketitle

\section{Introduction}

Terahertz time-domain spectroscopy (THz-TDS) has received much attention from researchers due to its outstanding performance~\cite{Mit99,Wit07c}. With the assistance of ultrafast femtosecond laser technology, a T-ray system generates and detects a short pulse, and thus the corresponding frequency range spanning from a few hundred gigahertz to a few terahertz or more. This frequency range has a relatively high black-body radiation background, prohibiting a high-SNR detection with a conventional detector. However, this problem is suppressed by adopting a coherent detection scheme, lifting the SNR to as high as 60~dB~\cite{Hua04}.
 
In the T-ray frequency region, many polar gases of general interest possess unique rotational transition energies, which give rise to spectral resonances \cite{Mit98}. Because of these unique fingerprints, THz-TDS proves useful for gas classification and recognition. However, this property, on the other hand, affects the THz-TDS capability in an open air setting, in which water vapor is ubiquitous. 

Water vapor, the third most abundant gas in nature~\cite{Tri03}, is known to have numerous rotational resonances in the T-ray band~\cite{Ext89}. THz-TDS of a sample, in open air, therefore unavoidably results in a combination of the sample's spectral features and water vapor resonances in the frequency domain. In the time domain, this results in field fluctuations after the main T-ray pulse. Mostly, these effects are undesirable, since they can obscure spectroscopic results of interest. 

The water-vapor effects are removable during the measurement by purging an enclosed T-ray path with dry air or such a non-polar gas as nitrogen, which does not have transition energy levels in the T-ray regime \cite{Gri90}---alternatively a vacuum is sometimes used. In some applications, it is not always possible to enclose the entire T-ray beam path, for example, in applications where stand-off detection is required \cite{Fed05}. The effects can be partially canceled out by calculation, if two measurements, one for the sample and the other for the reference, are available. Again, not all situations can provide both measurements. Performing narrowband T-ray sensing in an atmospheric transmission window could avoid the effects~\cite{Liu07}, at the expense of a full spectral fingerprint. As a completely different approach, a numerical method to alleviate the water vapor effects on the data would therefore be beneficial. 

Knowing the resonances' characteristics in the T-ray frequency range allows a numerical estimation of the complex response of water vapor. Theoretically, the estimated complex response can be deconvolved directly from a received T-ray signal. But pragmatically, fitting of a modeled complex response to a measured response is complicated by many factors, e.g. the limited dynamic range~\cite{Jep05}, limited frequency resolution~\cite{Xu03}, and measurement uncertainties \cite{Wit07d}. Besides, the exact model requires precise knowledge of geometric and atmospheric conditions during the measurement. Moreover, if direct deconvolution is carried out, there is no measure to validate the result.

Fine-tuning the strengths of complex resonances based on a brute-force search is introduced in this work. Each resonance is tuned in magnitude within a predefined range and then deconvolved from a measured signal. A tuning criterion is met when the ratio of the fluctuation energy to the main pulse energy of an adjusted signal is minimized. Repeatedly tuning the strength line-by-line ultimately results in the mitigation of water-vapor-induced effects. Constrained by the condition of minimum fluctuation energy, the algorithm always provides the optimum results.

The proposed algorithm based on signal processing has merits in that (i) its generality allows application to T-ray signals with reduced effect on samples' temporal and spectral features, and (ii) the exact measurement conditions, including humidity, temperature, pressure, and propagation length, are not crucial in order to remove the water-vapor response.

This paper is organized as follows: Section~\ref{sec:WV_vapor_effects} elaborates the effects of water vapor on T-ray signals and spectra. Section~\ref{sec:WV_model_abs_dis}, the model characteristics of each complex resonance is determined and validated with a measurement. In Section~\ref{sec:WV_line_tuning}, the algorithm for removal of water-vapor effects is introduced. This algorithm is verified by experiments in Section~\ref{sec:WV_results}, followed by discussions and a conclusion in Section~\ref{sec:WV_conclusion}.

\section{Effects of water vapor on pulsed T-ray signal}\label{sec:WV_vapor_effects}

\begin{figure}
	\centering
		\includegraphics[width=\columnwidth]{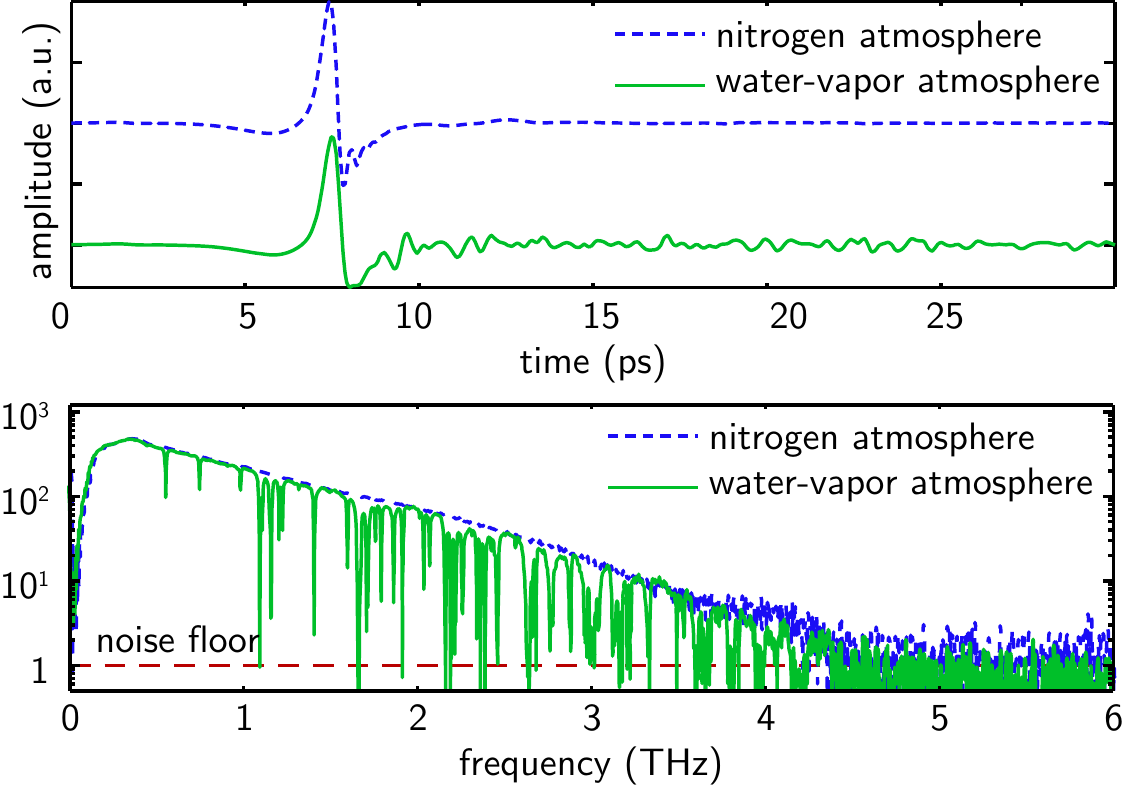}
	\caption{Effects of water vapor on a T-ray pulse and its spectrum. (above) T-ray signals recorded in nitrogen and water-vapor atmospheres. The sampling interval is 0.067~ps, and the total time duration is 136~ps. (below) Their corresponding spectra with the spectral resolution of 7.32~GHz. For this measurement the noise floor marks the cutoff frequency at approximately 4.0~THz.}
	\label{fig:WV_time_freq_signals}
\end{figure}

Water vapor contains many modes and the details of its interaction with an electromagnetic wave depends on the frequency. Water vapor exhibits pure rotational modes of energy transitions spanning the millimeter wave to mid-infrared or approximately 3~GHz to 19.5~THz \cite{Kem78,Pet02}. In the infrared region, both pure rotational line transitions and rotation-vibration bands are noticeable \cite{Gas88}. These transitions cause absorption and re-emission of wave energy in a number of narrow frequency bands, unique to water molecules.


Fig.~\ref{fig:WV_time_freq_signals} shows T-ray signals and spectra measured by THz-TDS under nitrogen and water-vapor atmospheres at room temperature and pressure. In the time domain, the signal recorded in the presence of water vapor undergoes long fluctuations after the main pulse. This is due to the energy re-emission by rotational transitions of water molecules. In the frequency domain, the water vapor causes several sharp resonances at discrete frequencies, as a result of quantised rotational transition energies. In terms of Fourier theory, the pair of time-domain fluctuations and frequency-domain resonances is based on the principle that a sharp feature in one domain is related to a broad feature in the other domain.

In addition to the fluctuation and the resonance effects, the analysis shows that the T-ray energy loss by water absorption calculated between 0.0 and 4.0~THz is as high as 10\% for the spectrum in Fig.~\ref{fig:WV_time_freq_signals}. The loss is likely due to the non-directional energy re-emission of water molecules. In the time domain, the ratio of the main pulse energy to the tail (fluctuation) energy is calculated for both of the time-domain signals (for more details on the calculation, see Section~\ref{subsec:WV_fluctuation_ratio}). The energy ratio for the nitrogen measurement is 429.98, whereas the ratio for the water vapor measurement is 18.82. This measure will be used later on to construct a criterion for fluctuation removal.


\section{Model of water vapor absorption and dispersion}\label{sec:WV_model_abs_dis}

It is required that the water-vapor response be modeled appropriately before a further step in water-vapor removal is taken. Modeling of absorption and dispersion for a rotational resonance involves selecting a suitable lineshape and calculating the line position, line strength, and linewidth. Fortunately, the line position and line strength are, only to a small extent, dependent on atmospheric conditions, i.e. the line position is constant at low pressures~\cite{Tow55} and the peak absorption is not a function of the pressure~\cite{Tow55,Har97}. In fact, the line position and line strength have been well parameterized in many publications. Thus these two parameters are readily available, leaving only the linewidth and lineshape to be determined. In the following subsections these parameters are discussed in detail relative to water molecule. 


\subsection{Line strength and line position}\label{subsec:WV_catalogues}

Water vapor rotational resonances in the T-ray regime have been extensively measured and analyzed via either conventional FTIR spectroscopy~\cite{Kau78,Mes83}, THz-TDS~\cite{Ext89,Che99}, or other techniques~\cite{Hel83,Mat95,Che00}---the extensive study is probably due to the abundance of water vapor, which involves many physical processes in nature. Subsequently, the computer-accessible spectroscopic parameters of water vapor, including the line position, line strengths and linewidth, are available from many research groups. Here in this work the catalog published by the JPL group~\cite{Pic98} is adopted. As this catalog has been regularly updated since before 1985~\cite{Poy85}, it is unlikely that the high-intensity lines of water vapor are absent from the list. 

According to the JPL catalog, the line strength is reported in terms of the integrated line intensity at 300~K, $I_{a}(300~\mathrm{K})$, and is scalable with the temperature, $T$. 
Despite the temperature dependency, the line strength is independent of pressure~\cite{Tow55}. Also included in the catalog is the line position. The shifting of the position is an order of magnitude lower than the line broadening~\cite{Pod04}, and thus the position is presumably pressure- and temperature-dependent~\cite{Tow55}.

\subsection{Linewidth}\label{subsec:WV_linewidth}


A rotational absorption resonance is possibly broadened by a number of factors, for instance, self or foreign-gas collisions, molecule-wall collisions, the Doppler effect, and natural lifetime broadening \cite{Tow55,Ber05}. Among these factors, at a standard pressure the collisional broadening is predominant in determining the spectral linewidth~\cite{Gop96,Har97}. 

More specifically, the collision broadened linewidth is dependent on pressure \cite{Ber05} and temperature~\cite{Gas88,Che99,Pod04}. This dependency yields the Benedict and Kaplan relationship~\cite{Ben64,Kem78},
\begin{eqnarray}
	\Delta\omega&=&\Delta\omega_0\left(\frac{p}{p_0}\right)\left(\frac{T_0}{T}\right)^{m}\;,
\end{eqnarray}
where $\Delta\omega_0$ is the FWHM of the line at pressure $p_0$ and temperature $T_0$, $\Delta\omega$ is the FWHM at pressure $p$ and temperature $T$, and $m$ is the temperature index, varying between 0.5 and 1.0. An index of $m=0.68$ can be used in the calculation~\cite{Rot05}.

Determined from the experiment, the average FWHM, $\Delta\omega_0/2\pi$, of water vapor resonance at ambient temperature and standard pressure is 6~GHz. This value is in agreement with an estimate of 10~MHz per Torr, or 7.6~GHz at 1 atm~\cite{Ber05,Luc03} within the limit of measurement uncertainty.

\subsection{Lineshape}\label{subsec:WV_lineshape}

The rotational absorption lineshape can be modeled either by Lorentz \cite{Lor06}, Vleck-Weisskopf \cite{Vle45}, or Gross~\cite{Eme72,Kem78} profiles. The choice depends on the frequency range of interest and the nature of line broadening. 
Notwithstanding, no significant difference is observable among these profiles at or nearby a T-ray resonance, with the linewidth nearly three orders of magnitude lower than the resonance frequency. In this paper, a Lorentzian profile is adopted in modeling the water vapor response in the T-ray frequency range.

A Lorentz absorption and dispersion profiles are given by~\cite{Kem78}, respectively,
\begin{subequations}\label{eq:WV_Lorentz_profile}
\begin{eqnarray}
\kappa_a(\omega)&=&\frac{c\Delta\omega_{a}}{2\pi}\left[\frac{1}{(\omega_{a}-\omega)^2+\Delta\omega_{a}^2}\right]\;,\\
n_a(\omega)-1&=&\frac{c}{2\pi}\left[\frac{\omega_{a}-\omega}{(\omega_{a}-\omega)^2+\Delta\omega_{a}^2}\right]\;,
\end{eqnarray}
\end{subequations}
where $\omega_a$ denotes the $a^{\rm th}$ transition frequency, and $\Delta\omega_{a}$ denotes the HWHM of the profile at that frequency. The absorption and dispersion profiles are linked together via Kramers-Kr\"onig relation.



\subsection{Ensemble of rotational transition resonances}\label{subsec:WV_ensemble_of_transitions}

This subsection gives a complete model, in terms of optical constants, as a result of several rotational transition resonances in the frequency range of interest. The model is a function of the lineshape, linewidth, line strength, and line position, which are discussed in the previous subsections. Later on, the model is compared to the measurement to verify the accuracy.

The model extinction coefficient and index of refraction are summation of an offset and a set of Lorentzian profiles, or, respectively,
\begin{subequations}
\begin{eqnarray}
	\kappa(\omega)&=&\kappa(0)+\sum_{a} m_{a}\kappa_{a}(\omega)\;,\\
	n(\omega)-1&=&n(0)+\sum_{a}m_{a}[n_{a}(\omega)-1]\;,
\end{eqnarray}\label{eq:WV_resonance_ensemble}
\end{subequations}
where the line strength $m_a$ is proportional to the integrated line intensity, $I_a$, given in the JPL catalog, divided by the line position, $\omega_a$, or, $m_a\propto I_a/\omega_a$. The offset is partially a result of electronic and molecular vibrational resonances at high frequencies~\cite{Kem78}, and also a contribution from the tails of other rotational resonances unaccounted for by the model. 

Typically, at the low frequency range, from DC to a few hundred gigahertz, THz-TDS cannot produce sufficient energy to overcome the noise. As a result, the resolved optical constants in this frequency range are unreliable. A phase extrapolation technique is usually introduced to correct the unwrapping process, forcing the phase to start at zero~\cite{Wit05a}. Thus, it is not necessary in the model to consider the index offset, $n(0)$, which is derived from the phase.

\begin{figure}
	\centering
		\includegraphics[width=\columnwidth]{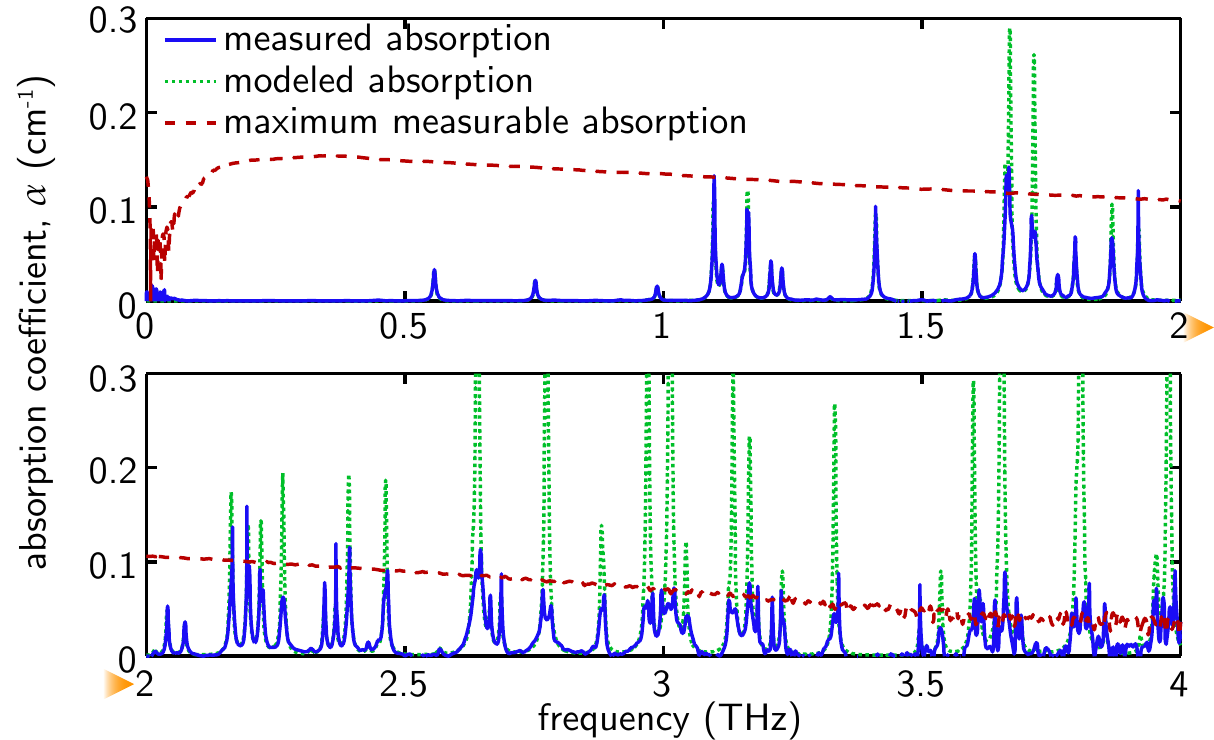}
	\caption{Model and measurement of H$_2$O absorption. The lower graph is a continuation of the upper graph, as indicated by the arrowheads. The measured and modeled curves match well in the range of 0.0 to 1.6~THz, but match poorly in the higher frequency range, where the absorption strengths become higher than the dynamic range of the system. The maximum measurable absorption coefficient of the system, represented by the dashed line, is calculated using the method of Jepsen and Fischer \cite{Jep05}.}
	\label{fig:WV_comparison_absorption}
\end{figure}


The absorption of water vapor, shown in  Fig.~\ref{fig:WV_comparison_absorption}, is determined from the measured data (Fig.~\ref{fig:WV_time_freq_signals}) and the Lorentzian model in the frequency range between 0.0 and 4.0~THz. It can be clearly seen that below 1.6~THz the model closely resembles the measurement. This match is possible since the resonances in this low frequency range have their strengths beneath the maximum absorption coefficient measurable by the system. However, as the frequency goes beyond 1.6~THz, the model cannot track the measurement due to the dynamic range of the system. In this situation, the system can no longer measure the absorption coefficient correctly, and the clipped absorption peaks are obvious, in particular, in the range from 2.0 to 4.0~THz. Furthermore, the clipped lines become less distinct, when the T-ray power reaches the noise floor beyond 3.0~THz. 

\subsection{Continuum absorption}\label{subsec:WV_continuum_absorption}

The continuum absorption, unaccounted for so far, is defined as the excess measured absorption unable to be quantified by the resonance spectrum. The mechanism underpinning the continuum has not been fully understood~\cite{Pod05}, and most of the mathematical models are fitted to experimental observations. 

It is known that the continuum absorption is pressure and frequency dependent. A continuum for H$_2$O collisions, derived from an experiment in the T-ray frequency range, between 0.4 and 1.83~THz, is estimated at $4.22\times 10^{-8}$ (dB/km)/(hPa~GHz)$^2$ \cite{Pod05}. A normal atmospheric measurement in a relatively short path length is insignificantly influenced by the continuum absorption \cite{Clo89,Pod05}.


\section{Removal of H$_2$O response by strength tuning}\label{sec:WV_line_tuning}

This section proposes a numerical means for removal of water-vapor effects, as an alternative to the conventional `sealed chamber' method. Subsection~\ref{subsec:WV_black_boxes} establishes the T-ray spectroscopic system, and points out the difficulties in using a direct deconvolution to remove the water vapor response. The strength-tuning algorithm is introduced in Subsection~\ref{subsec:WV_fine_tuning}, followed by the fluctuation ratio as a tuning criteria in Subsection~\ref{subsec:WV_fluctuation_ratio}. Discussion on the algorithm's generality can be found in Subsection~\ref{subsec:WV_generality}.

\subsection{Water vapor as a black box}\label{subsec:WV_black_boxes}

\begin{figure}[b]
	\centering
		\includegraphics[width=\columnwidth]{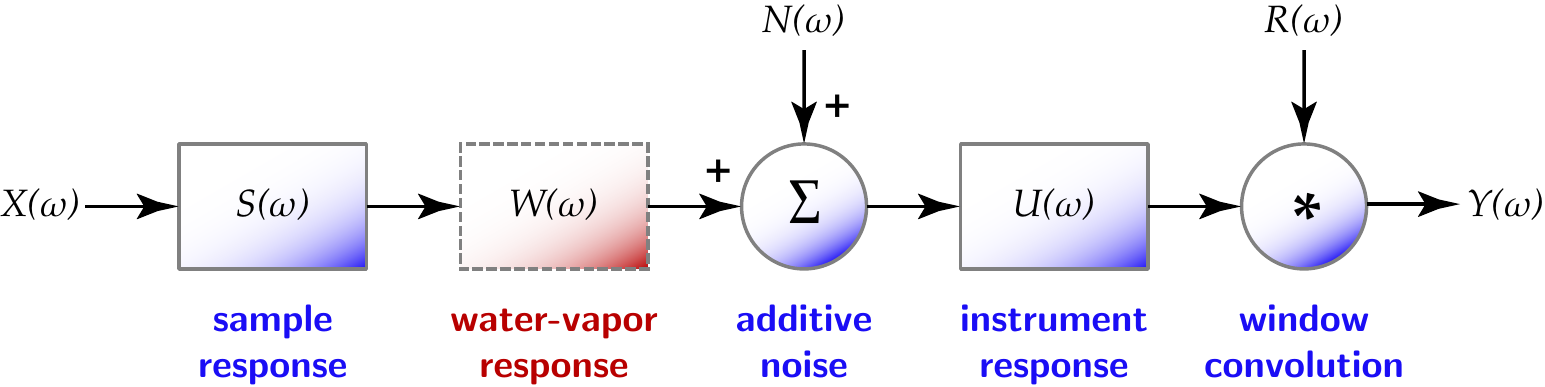}
	\caption{Model of a spectroscopic system. The system response is the combination of the frequency responses of the sample, water vapor, and instrument, denoted by $S(\omega)$, $W(\omega)$, and $U(\omega)$, respectively. The symbols, $X(\omega)$, $Y(\omega)$, $N(\omega)$, and $R(\omega)$ represent input T-ray spectrum, measured spectrum, noise, and window (or limited recording duration) responses, respectively. The water-vapor response in the dotted box needs to be removed.}
	\label{fig:WV_water_box}
\end{figure}

During a THz-TDS measurement, an emitted T-ray signal evolves based on several factors, e.g. the sample response, the water vapor response, the instrument response, and the noise. These can be modelled as a system, as illustrated in Fig.~\ref{fig:WV_water_box}. Given that $x(t)$ denotes the input T-ray pulse, which is immediately deployed from a transmitter, $s(t)$ the impulse response of a sample, $w(t)$ the impulse response of water vapor, $u(t)$ the impulse response of the instrument, $n(t)$ the noise, and $r(t)$ the time windowing, the received pulse, $y(t)$, can be described as
\begin{eqnarray}\label{eq:WV_td_system}
	y(t)&=&[x(t)\ast s(t)\ast w(t)\ast u(t)]\cdot r(t)\nonumber\\&&+[n(t)\ast u(t)]\cdot r(t)\;,
\end{eqnarray}
where $\ast$ is the convolution operator. The noise is simply additive and can thus be treated as an extra term in Equation~\ref{eq:WV_td_system}.

Via Fourier transform, the above equation in the frequency domain is
\begin{eqnarray}\label{eq:WV_spectro_system_freq}
	Y(\omega)=\left[X(\omega)\cdot S(\omega)\cdot W(\omega)\cdot U(\omega)\right]\ast R(\omega)+\acute{N}(\omega)\;,
\end{eqnarray}
where $Y(\omega)$, $X(\omega)$, $S(\omega)$, $W(\omega)$, $U(\omega)$, $R(\omega)$, and $\acute{N}(\omega)$ represent complex frequency responses, as the Fourier pairs of $y(t)$, $x(t)$, $s(t)$, $w(t)$, $u(t)$, $r(t)$, and $[n(t)\ast u(t)]\cdot r(t)$ respectively.

The water-vapor frequency response can be expressed by
\begin{eqnarray}\label{eq:WV_water_resp}
	W(\omega)&=&\exp\left[-j\hat{n}_{\rm vap}\frac{\rho(T)}{\rho_0(T)}\cdot\frac{\omega L}{c}\right]\;.
\end{eqnarray}
Here, $L$ denotes the T-ray propagation length in free space, excluding the space occupied by sample(s). The water vapor's complex index of refraction, $\hat{n}_{\rm vap}=n_{\rm vap}-j\kappa_{\rm vap}$, contains the refractive index $n_{\rm vap}$ and the extinction coefficient $\kappa_{\rm vap}$, which are modeled in Equation~\ref{eq:WV_resonance_ensemble}. The water-vapor response is also determined by the humidity~\cite{Yua03}, where $\rho(T)$ is the vapor density at temperature $T$, $\rho_0(T)$ is the saturation vapor density at the same temperature. 

In order to remove the effects of water vapor from a measurement, the water-vapor response, $W(\omega)$, in Equation~\ref{eq:WV_spectro_system_freq} must be replaced by the vacuum response, $V(\omega)$, with the same propagation length,
\begin{eqnarray}\label{eq:WV_vacuum_resp}
	V(\omega)&=&\exp\left[-j\omega L/c\right]\;.
\end{eqnarray}
From Equation~\ref{eq:WV_spectro_system_freq} the noisy response can be modelled in two parts,
\begin{eqnarray}\label{eq:WV_measured_spectrum_2_cases}
Y(\omega)\approx\left\{ \begin{array}{l}
\left[X(\omega)\!\cdot\! S(\omega)\!\cdot\! W(\omega)\!\cdot\! U(\omega)\right]\ast R(\omega)\\ \qquad\qquad\qquad\;\mathrm{, if}\quad |X\!\cdot\! S\!\cdot\! W|>|N|\\
\acute{N}(\omega) \qquad\qquad{\mathrm{, otherwise.}}\\
\end{array} \right.
\end{eqnarray}
Assuming the noise level is unity and the model water-vapor response matches well with the measurement, performing the direct deconvolution would yield
\begin{eqnarray}
	Y(\omega)\approx\left\{ \begin{array}{l}
\left[X(\omega)\cdot S(\omega)\cdot V(\omega)\cdot U(\omega)\right]\ast R(\omega) \\
\qquad\qquad\qquad\quad\;\mathrm{, if}\quad |X\cdot S\cdot W|>1\\
V(\omega)/W(\omega) \qquad\mathrm{, otherwise.}\\
\end{array} \right.
\end{eqnarray}
Therefore, removal of the water vapor response fails at \mbox{$|X\cdot S\cdot W|\leq |N|$} or $\mathrm{D}(\omega)\leq |S\cdot W|^{-1}$, where $\mathrm{D}(\omega)=|X/N|$ is the system's dynamic range. This usually occurs at the peaks of absorption resonances, as shown in Fig.~\ref{fig:WV_comparison_absorption}.

Consider the water vapor response closely. It can be separated into several components, and the separation is formulated as
\begin{eqnarray}\label{eq:WV_water_components}
	W(\omega)&=&V(\omega)W_0(\omega)W_1(\omega)\ldots W_a(\omega)\;.
\end{eqnarray}
Each component corresponds to a complex resonance, and has its frequency response derived from Equations~\ref{eq:WV_resonance_ensemble} and \ref{eq:WV_water_resp}, or
\begin{eqnarray}\label{eq:WV_repsonse_single_resonance}
	W_a(\omega)&=&\exp\left[-j\;m_a(n_a-j\kappa_a) \omega/c\right]\;.
\end{eqnarray}
The line strength, $m_a$, now incorporates the humidity ratio, $\rho(T)/\rho_0(T)$, and the propagation length, $L$. It can be inferred from Equations~\ref{eq:WV_water_components} and \ref{eq:WV_repsonse_single_resonance} that all components or resonances can be removed from the measured spectrum separately and independently. Hereby, the deconvolution at the noisy parts of the spectrum can be avoided, and also it is possible to tune $m_a$ of each resonance to the measurement without accurate information of the atmospheric and geometric parameters. In order to achieve this parameter relaxation, somehow, a criterion indicating the fit of tuning is essential. 


\subsection{Fluctuation-removal algorithm}\label{subsec:WV_fine_tuning}

\begin{figure}
	\centering
		\includegraphics[width=\columnwidth]{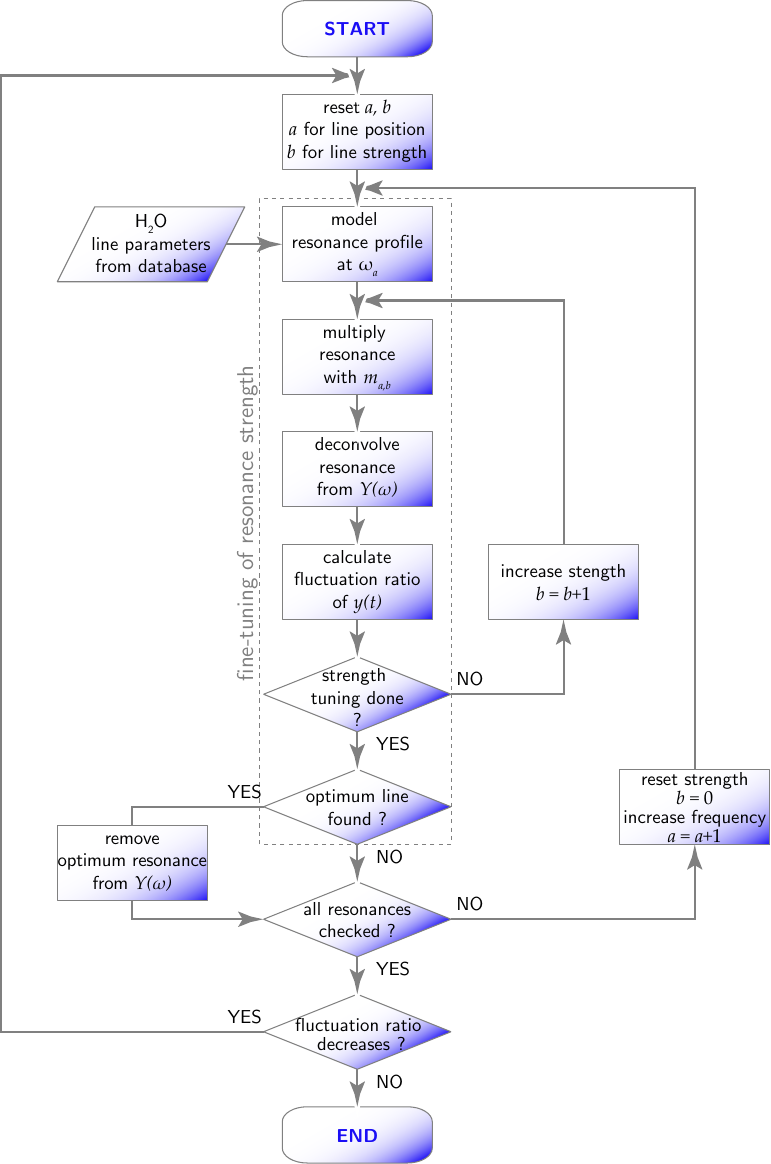}
	\caption{Fluctuation-removal algorithm. Each resonance in the T-ray frequency range is modeled and fine-tuned in its strength with the criterion of minimum fluctuation, and then the optimal resonance is removed from the measured complex response. The procedure is repeated until all resonances within the frequency range of interest are investigated, and until the fluctuation ratio no longer decreases.}
	\label{fig:WV_flowchart}
\end{figure}

The fluctuation-removal algorithm is shown in Fig.~\ref{fig:WV_flowchart}. The counters $a$ and $b$ for the line position and line strength are reset. The H$_2$O spectral line parameters, including intrinsic line positions and strengths, in the frequency range of interest, are then fetched from an existing database. A complex resonance profile, $n(\omega)-j\kappa(\omega)$, of a rotational transition at $\omega_{a=0}$ is modeled according to the Lorentzian profile in Equation~\ref{eq:WV_Lorentz_profile}. The modeled profile multiplied by the initial strength factor $m_{a=0,b=0}$ is then temporarily deconvolved from the measured complex response, $Y(\omega)$. The deconvolved time-domain signal, $y(t)$, is estimated for its fluctuation ratio (see Subsection~\ref{subsec:WV_fluctuation_ratio}). The procedure repeats to find the fluctuation ratio at other predefined line strengths $m_{a=0,b}$; $b\in\{1,2,3,\ldots\}$. 

Once the tuning range, $m_{a=0,b}$, is covered, the algorithm picks up an optimal strength within $m_{a=0,b}$, which gives the minimum fluctuation ratio, if any, and permanently removes that optimal complex resonance from the measured signal. The tuning procedure starts again, but with the next transition resonance, $\omega_{a=1}$, and is repeated until all resonances are optimized and possibly removed. It should be noted that a strong resonance, which suffers from noise i.e. $|X\cdot S\cdot W|<|N|$ at and around the peak of resonance, will be avoided by the tuning process, because performing deconvolution of such an ill-defined resonance will give rise to a high fluctuation ratio. Since there might be dependencies among the resonances, with regard to time-domain fluctuations, the whole process starts again until the fluctuation ratio no longer decreases. 

It is advisable not to perform zero padding prior to Fourier transform in the removal process, as the interpolated spectrum does not exactly reflect the reality---this is especially the case for points at resonances that have high variability. Otherwise, discrepancy between the model and the spectrum would cause a large remnant in the spectrum after deconvolution. Also note that to speed up the algorithm, the resonances that have a strength lower than the amplitude uncertainty can be skipped.

\subsection{Fluctuation ratio}\label{subsec:WV_fluctuation_ratio}

A criterion related to the quality of a T-ray signal is necessary in selecting the optimal resonance line strengths. The `quality' here is defined as having a high pulse energy with low fluctuations in the time domain, which implies the absence of water-vapor resonances in the frequency domain. 

In order to quantify the quality one must be able to evaluate the total energies of the main pulse and of the fluctuations. One potential way is to window the time-domain signal with a Gaussian profile to obtain the amplitude at a desired portion. A Gaussian window is selected, because a T-ray main pulse is essentially derived from an optical pumping pulse, which takes on a Gaussian pulse profile \cite{Duv01}.

\begin{figure}
	\centering
		\includegraphics[width=\columnwidth]{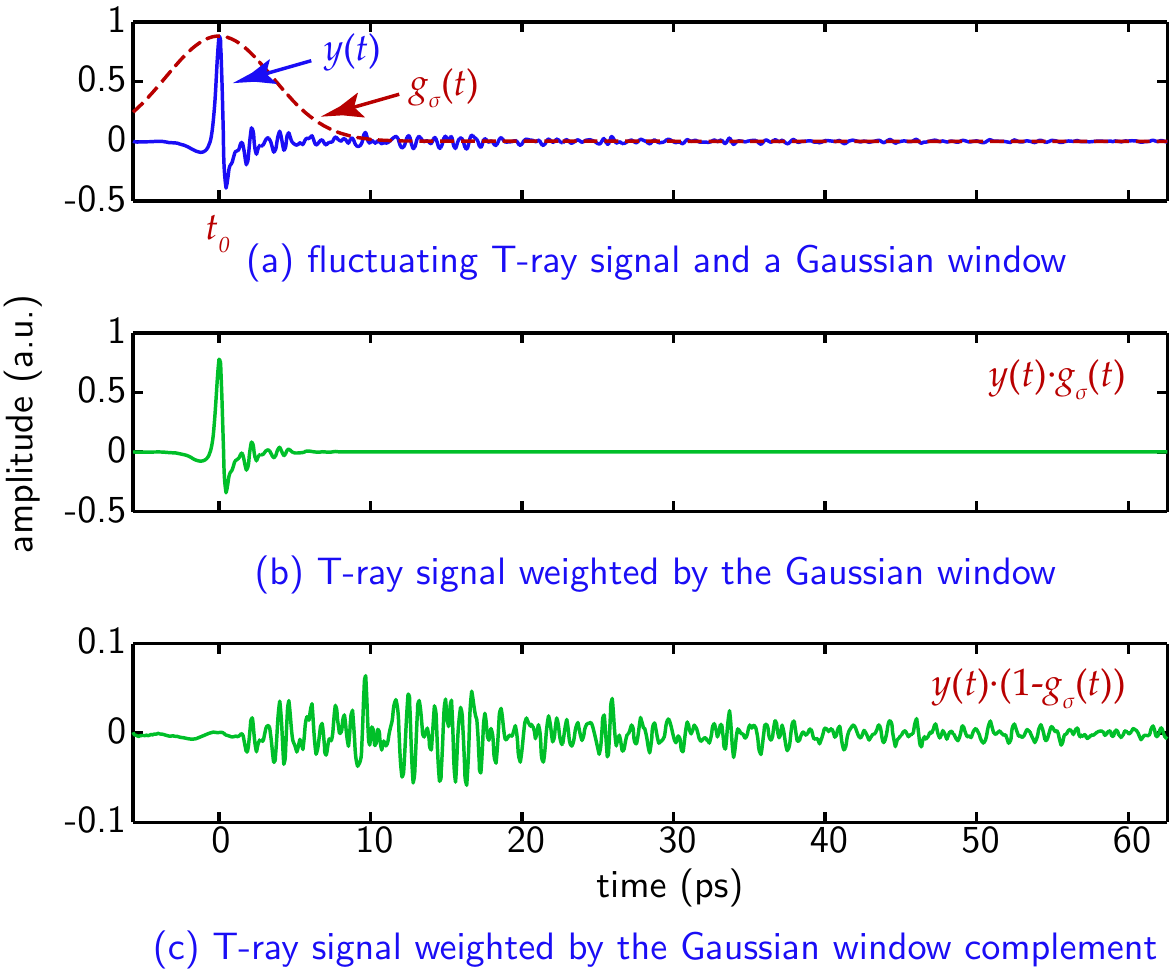}
	\caption{Gaussian window applied to the T-ray signal. A T-ray signal weighted by a Gaussian window, which has its peak position set exactly at the main pulse peak, yields the main pulse with suppressed fluctuations. A T-ray pulse weighted by the Gaussian window complement yields only fluctuations. The FWHM of the Gaussian window used here is 8.33~ps.}
	\label{fig:WV_window_gaussian}
\end{figure}

Fig.~\ref{fig:WV_window_gaussian}(a) shows a Gaussian window, $g_\sigma(t)$, overlapping the T-ray signal, $y(t)$. A Gaussian window with appropriate width and position would eliminate the fluctuating tail of a signal, as shown in Fig.~\ref{fig:WV_window_gaussian}(b). On the other hand, the complement of a Gaussian window could also be used to remove the main pulse, as in Fig.~\ref{fig:WV_window_gaussian}(c). With the assistance of a Gaussian window, the fluctuation energy normalised by the main pulse's total energy---coined the \emph{fluctuation ratio}---can be formulated as
\begin{eqnarray}\label{eq:WV_fluctuation_ratio}
	F(y,g\;;\sigma)&=&\frac{\int_t[y(t)\cdot (1-g_\sigma(t))]^2dt}{\int_t[y(t)\cdot g_\sigma(t)]^2dt}\;.
\end{eqnarray}
The integration is carried out over the time duration of a recorded T-ray signal.
A Gaussian window is given by
\begin{eqnarray}\label{eq:WV_gaussian_window}
	g_\sigma(t)&=&\exp[-(t-t_0)^2/2\sigma^2]\;,
\end{eqnarray}
where $t_0$ is the peak position of the T-ray signal, $y(t)$, and $\sigma$ multiplied by $2\sqrt{2\ln 2}$ is the FWHM of a Gaussian window.



\subsection{Generality of the algorithm}\label{subsec:WV_generality}

The algorithm is plausibly general---in the sense that it can be applied to any T-ray signal with minimal disturbance of desired signal features. This generality is due to the following facts: 

First of all, a T-ray spectrum is altered by the algorithm only the frequencies at which water-vapor absorption lines are situated. The spectral resonances of other polar gases, occupying narrow frequency bands and usually not overlapping with water resonances, are not disturbed. In addition, the broad spectral features of a time-domain transient is unlikely to be significantly affected by narrow water line removal. The reflections, for example, which exhibit fringes over a broad frequency range, are not affected by narrow band resonance removal.

Second, unrelated fluctuations are ignored. Because the algorithm senses the fluctuation change, caused by strength tuning, rather than the fluctuation itself, any unrelated fluctuations cannot deceive the algorithm.

Last, the causality of a signal is preserved. The absorption and dispersion profiles, used in the model of water vapor response, exactly comply with the Kramers-Kr\"onig relation. Deconvolution of the measured spectrum by this causal response would yield the causal spectrum.

\section{Results}\label{sec:WV_results}

\subsection{Free-path measurement}

The first T-ray signal, subject to the water-vapor removal algorithm, is measured in free space without the presence of any material. The signal has the temporal resolution of 66.7~fs and the total duration of 136.63~ps, providing the spectrum with the spectral resolution of 7.3~GHz.

The strength-tuning algorithm is carried out with a set of complex resonances in the frequency range between 0 and 4~THz, where the T-ray magnitude is relatively high. The sequence, in which the resonances are interrogated and removed, follows the numerical order, i.e. from low to high frequencies. Only the resonances that have the strength higher than 0.01 of the maximum strength are inspected. The FWHM of the Gaussian window is 3~ps, and the iteration of the algorithm is set to five.

\begin{figure}
	\centering
		\includegraphics[width=\columnwidth]{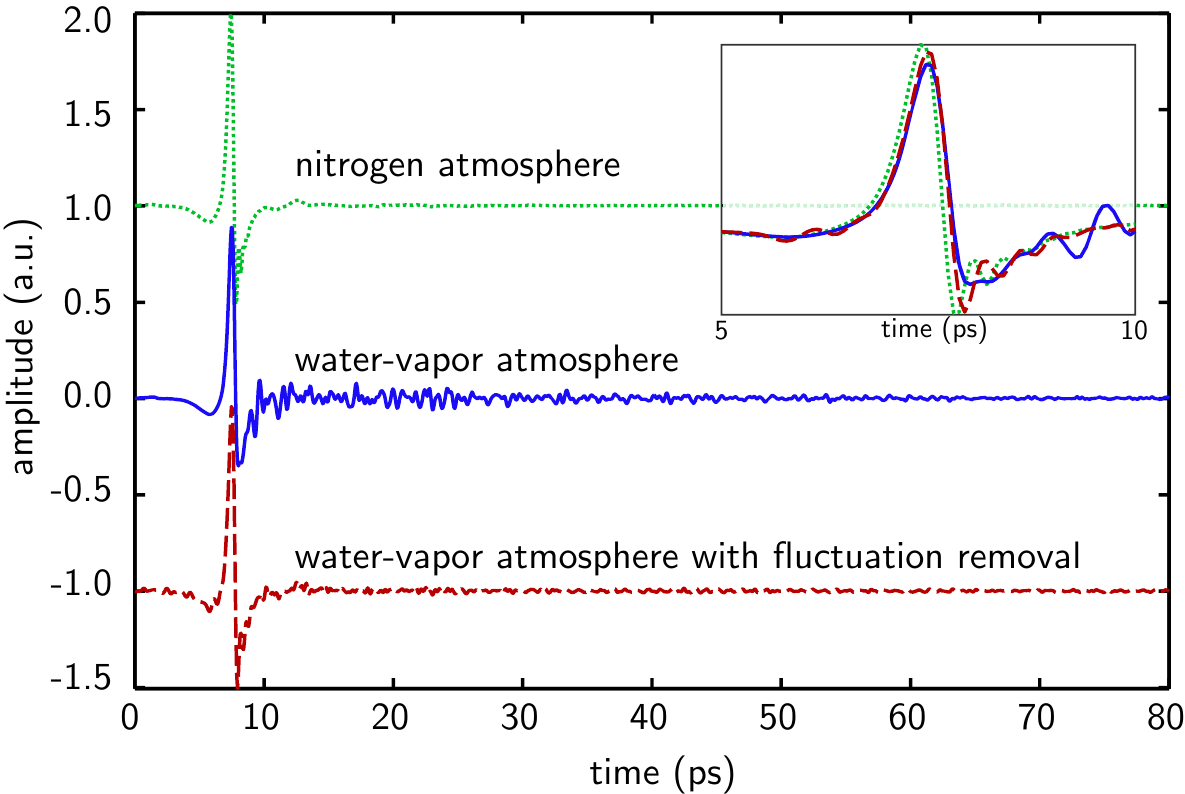}
	\caption{T-ray signals measured in free path. The sampling interval is 66.7~fs with the total duration of 136.63~ps (only the first 80~ps is shown here). The inset shows a zoom-in of signals between 5 and 10~ps. The fluctuation ratios of the N$_2$-atmosphere, H$_2$O-atmosphere, and H$_2$O-removal measurements are 2.2, 9.2, and 3.3, respectively. The fluctuation after the main pulse is remarkably reduced.}
	\label{fig:WV_fluc_removed_freepath}
\end{figure}

\begin{figure}[b]
	\centering
		\includegraphics[width=\columnwidth]{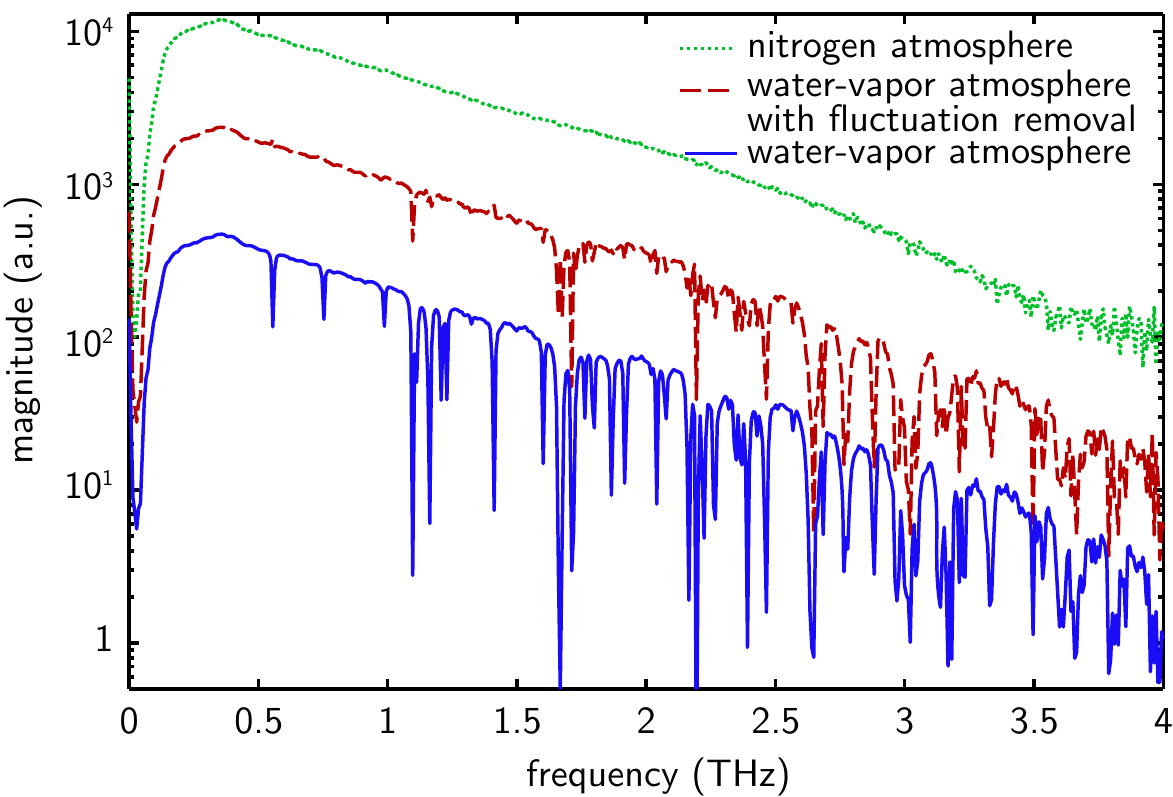}
	\caption{T-ray spectra measured in free path. The spectral resolution is 7.3~GHz. Up to 2.5~THz most of the water resonances are removed, or at least reduced in the strength, by the proposed algorithm. The magnitudes are offset for clarity.}
	\label{fig:WV_fluc_removed_freepath_spectrum}
\end{figure}

As shown in Fig.~\ref{fig:WV_fluc_removed_freepath}, the algorithm can significantly reduce the time-domain fluctuations, which are previously located immediately after the main pulse, i.e. after 10~ps. The main pulses in the inset	clearly illustrate the similarity between the fluctuation-free signal and the processed signal, despite a temporal shift which is not accounted for by the algorithm. In Fig.~\ref{fig:WV_fluc_removed_freepath_spectrum}, the spectrum result demonstrates success of the algorithm in removing, or reducing the strength of, the resonances in the frequency range from 0 to 2.5~THz. However, some resonances still persist, where they are spectrally congested or severely disturbed by noise.

\subsection{DL-Phenylalanine measurement}

The second experiment is performed with a signal measured from a DL-Phenylalanine sample \cite{Yam05} at 220~K. The cryostat containing the sample is evacuated, so no water vapor at cryogenic temperature involves in the measurement. The interacting water vapor is from normal air, surrounding the cryostat. The total scan is 68.3~ps at the step size of 66.7~fs. The spectral resolution is 14.6~GHz.

The recorded signal particularly contains the reflections, caused by the cryostat's windows, made from a cyclo-olefin copolymer (Topas). Use of such a signal demonstrates how well the proposed algorithm can deal with water-vapor fluctuations without disturbing unrelated reflection features. 

The parameters set for the algorithm are similar to the previous test, except for the interrogated frequency range which is in between 0 and 2~THz.

\begin{figure}
	\centering
	  \includegraphics[width=\columnwidth]{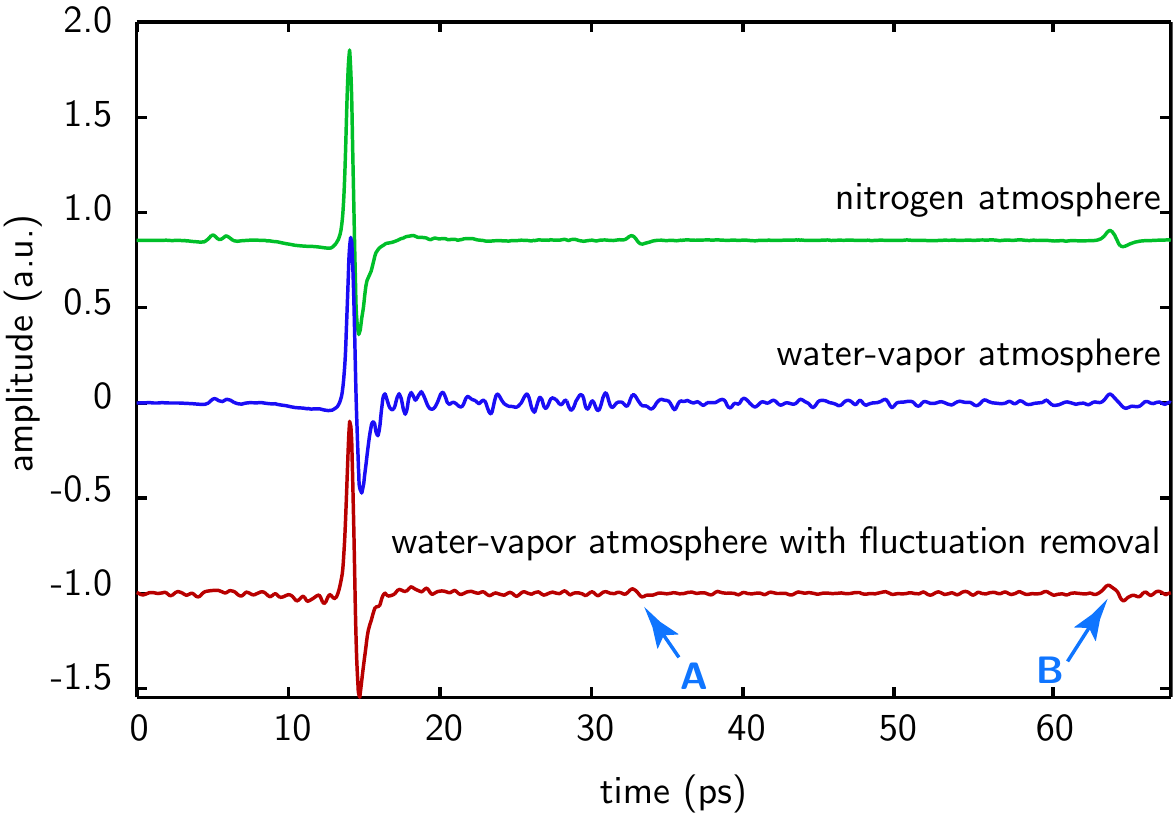}
	\caption{T-ray signals measured with DL-Phenylalanine sample in place. The sampling interval is 66.7~fs with the total duration of 68.38~ps . The reflections at 32 and 63~ps due to the cryostat's windows can be recovered from fluctuation by the algorithm. The fluctuation ratios of the N$_2$-atmosphere, H$_2$O-atmosphere, and H$_2$O-removal measurements are 2.6, 7.9, and 3.8, respectively.}
	\label{fig:WV_fluc_removed_alanine}
\end{figure}

\begin{figure}
	\centering
		\includegraphics[width=\columnwidth]{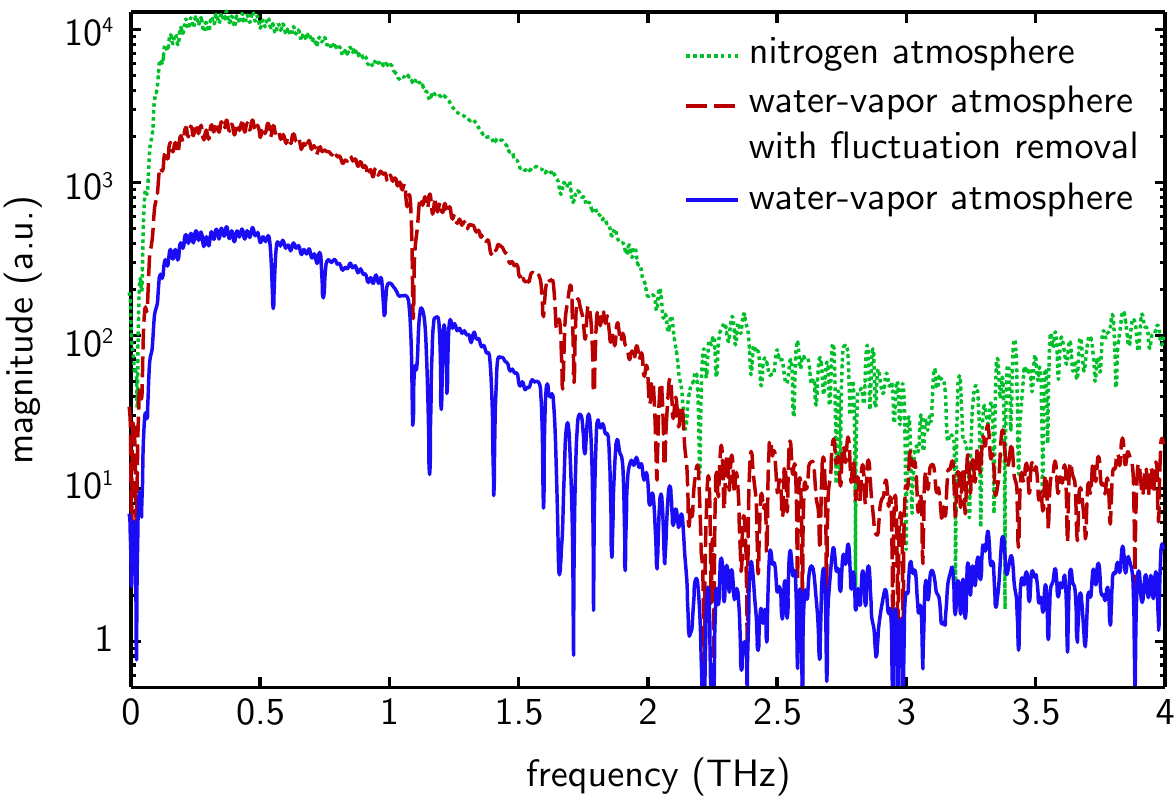}
	\caption{T-ray spectra measured with DL-Phenylalanine sample in place. The spectral resolution is 14.6~GHz. From 0 to 2~THz most of the resonances are removed, or at least reduced in the strength, by the proposed algorithm. The magnitudes are offset for clarity.}
	\label{fig:WV_fluc_removed_alanine_spectrum}
\end{figure}

The processed signal is shown in Fig.~\ref{fig:WV_fluc_removed_alanine}, along with the original. Again, it is obvious that the fluctuations located after the main pulse, beyond 15~ps, are remarkably reduced, in spite of an introduction of small oscillation at the beginning. The reflection at 63~ps (point B) is not disturbed by the algorithm, and, surprisingly, the reflection at 32~ps (point A), which was initially buried in fluctuations, is now recovered. In Fig.~\ref{fig:WV_fluc_removed_alanine_spectrum}, the algorithm can remove most of the H$_2$O resonances. Still a few resonances cannot be completely removed. These persistent resonances are around 1.1 and 1.7~THz, exactly the same positions as the unremoved resonances in the previous case (see Fig.~\ref{fig:WV_fluc_removed_freepath_spectrum}). The recovery of reflections suggests a promising application of the algorithm in T-ray range finding.

\subsection{Evaluation of the algorithm's performance}

Table~\ref{tab:WV_evaluation} shows the fluctuation ratio and mean squared error (MSE) of the signals and the total energy of the spectra, investigated in the previous subsections. Two additional results, not shown in this paper, are tested with lactose's reference and sample signals. The fluctuation ratio helps measure a change in the fluctuation energy and main pulse energy, before and after the signals are processed. The MSE evaluates the fitness of the improved signals to the N$_2$-atmosphere measurement. The MSE is given by
\begin{eqnarray}\label{eq:WV_MSE}
	\mathrm{MSE}&=&\frac{1}{m}\sum_m (\acute{y}_m-y_m)^2\;,
\end{eqnarray}
where $y$ is a signal measured in the nitrogen atmosphere and $\acute{y}$ is a compared signal. The error is summed and averaged over $m$ temporal points. The total spectrum energy shows the recovery of the energy from resonance absorptions. Note that the fluctuation ratio and the spectral energy are normalised to those of the signal measured in nitrogen atmosphere.

From the table the fluctuation ratio indicates that the fluctuation energy is greatly reduced in all cases, once the removal algorithm is implemented with the signal measured in the H$_2$O path. However, the MSE still reflects a considerable difference between the processed signals and the signals measured in a nitrogen-filled chamber, in spite of the great improvement visualized in the previous subsections. This would probably be due to the whole signal shift in the time domain, caused by a constant refractive index difference unaccounted for by the algorithm. 

\begin{table}
	\caption{Quantitative evaluation of the algorithm's performance. The notation $\mathrm{H}_2\mathrm{O}$, and $\overline{\mathrm{H}_2\mathrm{O}}$ stand for measurements in water vapor and water vapor with fluctuation removal, respectively.}
	\centering
  \begin{tabular*}{\columnwidth}
     {@{\extracolsep{\fill}}ccccccc}
\toprule 
\multirow{2}{*}{Signal}& \multicolumn{2}{c}{Fluctuation ratio}& \multicolumn{2}{c}{MSE (\%)}& \multicolumn{2}{c}{Energy (a.u.)}\\\cline{2-3}\cline{4-5}\cline{6-7}
&\tiny{$\mathrm{H}_2\mathrm{O}$}&\tiny{$\overline{\mathrm{H}_2\mathrm{O}}$}&
\tiny{$\mathrm{H}_2\mathrm{O}$}&\tiny{$\overline{\mathrm{H}_2\mathrm{O}}$}&
\tiny{$\mathrm{H}_2\mathrm{O}$}&\tiny{$\overline{\mathrm{H}_2\mathrm{O}}$}\\
\midrule
Free path&4.2&1.5&0.076&0.047&0.90&0.96\\
Phenylalanine&3.0&1.5&0.13&0.085&0.90&0.95\\
Lactose reference&16.1&2.0&0.31&0.19&0.90&0.98\\
Lactose sample&4.8&1.7&0.31&0.21&0.93&0.99\\
\bottomrule
\end{tabular*}	\label{tab:WV_evaluation}
\end{table}

The total spectral energy in Table~\ref{tab:WV_evaluation} reveals the algorithm is successful in recovering a part of the signal's energy, which is absorbed by water vapor resonances during propagation. However, not all the energy is recoverable. With regard to the processed spectra shown in the previous subsections, the algorithm cannot completely remove the water vapor resonances in the region where the SNR is low and where the spectral lines are crowded. The causes of these two cases are analyzed separately as follows.

At a low SNR region, distortions in the absorption and dispersion resonances by noise causes deviation of the measurement from the model. This situation is demonstrated in Fig.~\ref{fig:WV_comparison_absorption}. Applying the removal algorithm to a low SNR region would be effective, provided that the distortion is not severe. Otherwise, the large deviation inhibits the removal via a sub-optimal value of the fluctuation ratio. A possible means to alleviate the noise issue is to implement a signal enhancement methodology such as wavelet denoising~\cite{Fer01a} prior to the removal process; this might better recover water resonances and subsequently improve the removal efficiency.

Crowded resonance features result in overlapping and merging between two or more resonances. If the spectral resolution is too low, these blended resonances are rendered incorrectly, resulting in distorted shape. The algorithm would see them as a single stronger and wider resonance, and as a result, the removal does not perform well. Improving the spectral resolution would allow better removal performance. For example, the free space measurement in Fig.~\ref{fig:WV_fluc_removed_freepath_spectrum} shows a better result than the DL-Phenylalanine measurement in Fig.~\ref{fig:WV_fluc_removed_alanine_spectrum} in the regions around 1.1 and 1.7~THz; the former and latter measurements have the spectral resolutions of 7.3~GHz and 14.6~GHz, respectively. In addition to the spectral resolution problem, blended resonances cannot be represented exactly by a simple sum between two Lorentzian lines~\cite{Pic80}.

\section{Conclusion}\label{sec:WV_conclusion}

A numerical algorithm for removal of water-vapor effects in T-ray measurements is investigated in this paper. Given only a sample signal with no reference, theoretically, the removal can possibly be carried out by simple deconvolution of the model complex water-vapor response from the signal. However, many factors limit the usability of the simple deconvolution. An exact model of water vapor resonance requires many physical data, including the pressure, temperature, humidity, and propagation length. In addition, if the noise level is sufficiently high such that some strong resonances are distorted, the whole deconvolution process cannot be performed.

The proposed algorithm fine-tunes the strength of each model spectral resonance. A criterion for strength tuning is met when the fluctuation ratio of the processed signal reaches the minimum. Once an optimal resonance is attained, it is removed from the signal. The algorithm then proceeds to removal of the next resonance. This tuning scheme relaxes the requirement for precise information of the delicate physical conditions during the measurement. Furthermore, the fluctuation ratio criterion inhibits any fault deconvolution that might occur when the noise disturbs the quality of a measured signal.

In the experiments, the algorithm produces promising results. It can reduce a significant amount of the signal fluctuations and spectral resonances with small disturbances to other non-related features, such as reflections or sample-induced resonances. Moreover, some features, which are initially masked by fluctuations, can be recovered by the algorithm. Optical constants extracted from improved signals have an acceptable quality.

The proposed algorithm is general in the sense that any other polar gas response could be targeted, in principle, and hence removed from a measured spectrum if desired. It is not necessary that the parameters of a targeted gas be present in a spectroscopic catalog, as long as its pure response in the frequency range of interest is fully estimable. This scheme might have benefits in some particular situations, for instance, where molecular contaminant(s) are unavoidable in measurement.

Finally, it should be noted that the algorithm is not meant to serve as a full replacement to the `sealed chamber' procedure executed during the measurement stage. In fact, it assists the measurement in cases where a sealed chamber is not feasible. Still the algorithm cannot remove some traces of resonances due to the ubiquitous noise and resonance crowding. Further improvements of the algorithm remains an attractive challenge.

\section*{Acknowledgment}

The authors gratefully acknowledge Morten Franz at the Department of Molecular and Optical Physics, the University of Freiburg, for his support in T-ray measurements, and Brian W.-H. Ng at the School of Electrical \& Electronic Engineering, the University of Adelaide, for useful technical discussion. 

\bibliographystyle{IEEEtran}

\begin{biography}[{\includegraphics[width=1in,height=1.25in,clip,keepaspectratio]{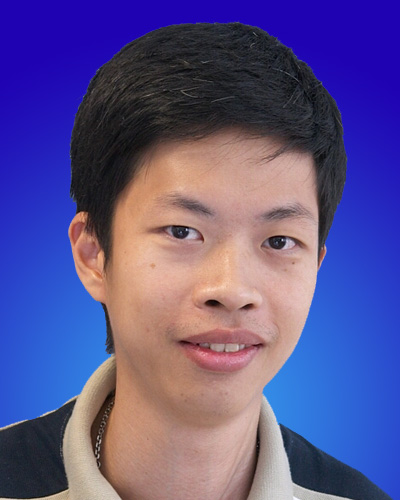}}]{Withawat Withayachumnankul}
was born in Bangkok, Thailand on December 20, 1980. From 1996-2000, he was educated at King Mongkut's Institute of Technology at Ladkrabang (KMITL), in his hometown, where he obtained a Bachelor's Degree in Electronics Engineering with honours. Working at the Biomedical Signal \& Image Processing Laboratory at KMITL in 2000, as a postgraduate student, he performed multi-disciplinary research involving digital signal processing, medical image reconstruction, and computer graphics. Here, he played a major role in developing core programming codes for medical software currently deployed in many leading Thai institutions and hospitals. After receiving a Master's degree in Electronics Engineering in 2002, he served at his alma mater as a lecturer, a position that he currently retains. In this role, he conducted the computer programming and computer graphics courses in the Department of Information Engineering. 

In 2005, he was a visiting scholar at the University of Adelaide, under Derek Abbott. In 2006, Withawat Withayachumnankul was granted an Australian Endeavour International Postgraduate Research Scholarship (EIPRS) to study toward his PhD under Derek Abbott, Bernd Fischer, and Samuel Mickan, within the Adelaide T-ray group, the School of Electrical \& Electronic Engineering, The University of Adelaide. Mr Withayachumnankul's current research involves T-ray signal processing, T-ray components, and superluminal (group delay) electromagnetic propagation.
\end{biography}

\begin{biography}[{\includegraphics[width=1in,height=1.25in,clip,keepaspectratio]{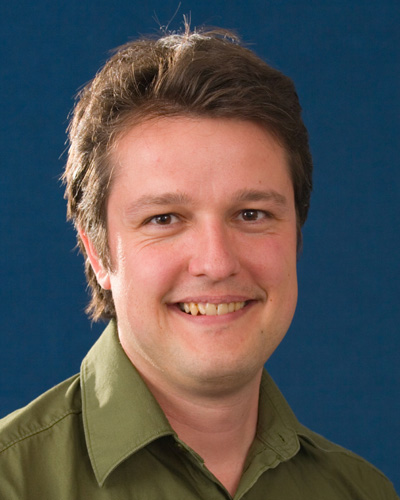}}]{Bernd M. Fischer} was born in Waldkirch, Germany. In 1997/98 he won an ERASMUS bursary and studied for one year in the Universit\'e Paris-sud 11, France, where he completed a {\it Stage de Maitrise} in biophysics and he received his Diplom Degree (with distinction) in 2001 from the University of Freiburg, Germany. He received his PhD degree (summa cum laude) in physics from the University of Freiburg, Germany, in 2006, under Peter Uhd Jepsen and Hanspeter Helm in the area of T-ray spectroscopy of biomolecules. 

Dr Fischer joined the Adelaide T-ray Group, Australia, in 2006, and in 2007 he was awarded the prestigious Australian Research Council (ARC) Postdoctoral (APD) Fellowship.
\end{biography}

\begin{biography}[{\includegraphics[width=1in,height=1.25in,clip,keepaspectratio]{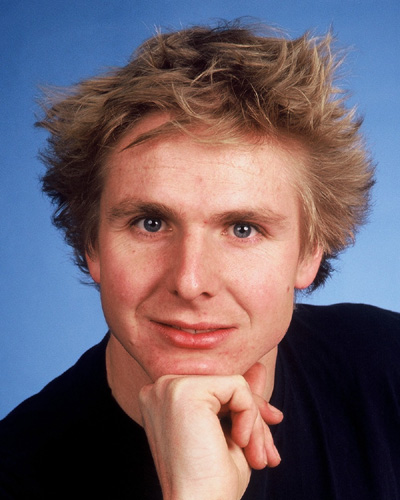}}]{Samuel P. Mickan}(S'99-M'03) was born on March 26, 1976 in Adelaide, Australia. He graduated with a BEng in Electrical and Electronic Engineering, (1997), and a with BA in German \& History (2002), both at the University of Adelaide. He received his PhD (with Special Commendation) in 2004 under Derek Abbott and Jesper Munch at the University of Adelaide, in the area of T-ray biosensing. He has won a number of awards and honors including, the Schneider Award for Excellence (1997), AFUW Barbara Crase Bursary (2001),  George Murray Scholarship (2002), the Australian Academy of Sciences (AAS) Young Researcher's Award (2004), the University Medal from the University of Adelaide (2004), and the Tall Poppy Award for science (2005). Under a Fulbright Scholarship, he collaborated with X.-C. Zhang's group, Center for Terahertz Technology, Rensselaer Polytechnic Institute in New York, USA, from 2001 to 2002. In 2003, he began lecturing in the School of Electrical and Electronic Engineering at the University of Adelaide. He is currently both an adjunct senior lecturer at the University of Adelaide and a patent attorney at Davies Collison Cave, Melbourne, Australia. He was Treasurer/Secretary of Microwave Theory \& Techniques Chapter of LEOS and was founder of the IEEE student branch at the University of Adelaide. Dr Mickan is a member of the IEEE, IEEE LEOS society, OSA, APS, and SPIE.
\end{biography}

\begin{biography}[{\includegraphics[width=1in,height=1.25in,clip,keepaspectratio]{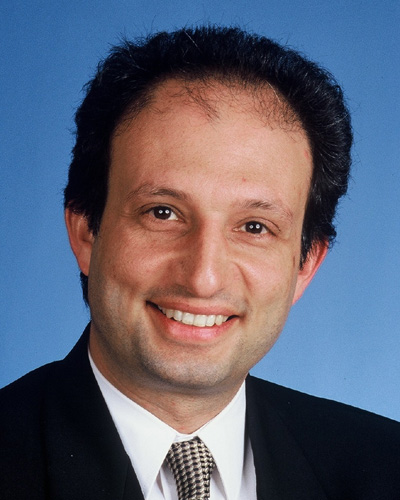}}]{Derek Abbott}(M'85-SM'99-F'05) was born on May 3, 1960, in South Kensington, London, UK, and he received a BSc (Hons) in physics (1982) from Loughborough University of Technology, UK. He completed his PhD (with commendation), in electrical \& electronic engineering (1995) from The University of Adelaide, Australia, under Kamran Eshraghian and Bruce~R.~Davis. He has led a number of research programs in the imaging arena, ranging from the optical to infrared to millimeter wave to T-ray (terahertz gap) regimes. From 1978 to 1986, he worked at the GEC Hirst Research Centre, London, UK, in the area of visible and infrared image sensors. His expertise also spans VLSI design, optoelectronics, device physics, and noise; where he has worked with nMOS, CMOS, SOS, CCD, GaAs, and vacuum microelectronic technologies. On migration to Australia, he worked for Austek Microsystems, Technology Park, South Australia, in 1986. Since 1987, he has been with The University of Adelaide, where he is presently a full Professor in the School of Electrical \& Electronic Engineering and the Director of the Centre for Biomedical Engineering (CBME). He has appeared on national and international television and radio and has also received scientific reportage in \textit{New Scientist}, \textit{The Sciences}, \textit{Scientific American}, \textit{Nature}, \textit{The New York Times}, and \textit{Sciences et Avenir}. He holds over 300 publications/patents and has been an invited speaker at over 80 institutions, including Princeton, NJ; MIT, MA; Santa Fe Institute, NM; Los Alamos National Laboratories, NM; Cambridge, UK; and EPFL, Lausanne, Switzerland. He won the GEC Bursary (1977), the Stephen Cole the Elder Prize (1998), the E.R.H.~Tiekink Memorial Award (2002), SPIE Scholarship Award for Optical Engineering and Science (2003), the South Australian Tall Poppy Award for Science (2004) and the Premier's SA Great Award in Science and Technology for outstanding contributions to South Australia (2004). He has served as an editor and/or guest editor for a number of journals including \textsc{IEEE Journal of Solid-State Circuits}, \textit{Chaos} (AIP), \textit{Smart Structures and Materials} (IOP), \textit{Journal of Optics B} (IOP), \textit{Microelectronics Journal} (Elsevier), \textit{Fluctuation Noise Letters} (World Scientific), and is currently on the Editorial Board of \textsc{Proceedings of the IEEE}. He has served on a number of IEEE technical program committees, including the \textit{IEEE APCCS} and the \textit{IEEE GaAs IC Symposium}. Prof Abbott is a Fellow of the Institute of Physics (IOP), with honorary life membership, and is a Fellow of the IEEE.
\end{biography}

\end{document}